\documentclass{svproc}
\usepackage{url}

\usepackage{graphicx}
\usepackage{subfigure}
\usepackage{amsmath}
\usepackage[braket, qm]{qcircuit}
\usepackage{arydshln}

\usepackage[none]{hyphenat}

\begin{document}
\mainmatter              
\title{QuantumSolver:\\A quantum tool-set for developers}
\titlerunning{QuantumSolver}  
%
\author{Daniel Escánez-Expósito\inst{1} \and Pino Caballero-Gil\inst{2} \and \\
Francisco Martín-Fernández\inst{3}}
\authorrunning{Escánez-Expósito, D.; Caballero-Gil, P.; Martin-Fernandez F.} 
%
\tocauthor{Daniel Escanez-Exposito, Pino Caballero-Gil and Francisco Martin-Fernandez}
\institute{University of La Laguna, Tenerife, Spain,\\
\email{jdanielescanez@gmail.com}
\and
University of La Laguna, Tenerife, Spain,\\
\email{pcaballe@ull.edu.es}
\and
IBM Research, New York, USA,\\
\email{paco@ibm.com}}

\maketitle              

\begin{abstract}
This paper introduces a new {\it opensource} quantum {\it tool-set} called {\it QuantumSolver} based on {\it Qiskit} to help developers without knowledge in quantum computing. The developed library includes a set of algorithms with different features: random number generation, Bernstein-Vazirani algorithm and quantum key distribution using the BB84 protocol. This paper described the main details about the implementation of the {\it toolset}, focusing in the challenges that the authors faced. Finally, this document analyzes the results obtained with some conclusions that authors compares with the included features.
\end{abstract}
\section{Introduction}

The interest in technologies related to quantum computing has been growing progressive in the last years, reaching the high peak nowadays. Its feature to hack the current classical cryptography algorithms has raised a real risk, triggering the dire need of new secure protocols for communications implemented from scratch. In the other hand, analyzing its magic properties, it is clear that quantum computing will be the next big thing in computation to speed up the human evolution from a technology point of view, generating solutios to problems never seen before \cite{GoogleQuantumSupremacy}.

The development of a quantum computing {\it open source tool-set} intended for developers, with a public repository in GitHub\cite{QuantumSolver}, pursue the abstraction and simple encapsulation of a quantum {\it software} with different features. In fact, promoting quantum computing technologies applied to software applications in the main scope of the proposal described in this scientific article.

{\it QuantumSolver} is oriented to users without experience in computer science (e.g. a user who wants to obtain a real random number generated thanks to quantum computing properties); but also is designed for senior developers that they want to support the library adding new methods or improving the current ones. It has been developed a specific mode to run the methods using two kind of user interfaces: {\it  Command Line Interface} (CLI) and Web Interface designed by the user experience.

\section{Implementation}

{\it QuantumSolver} is a quantum library developed in Python3 thanks to {\it Qiskit} (a open source SDK offered by IBM to play with quantum computing to different levels: pulses, circuits, applications... \cite{Qiskit}). {\it QuantumSolver} has two main components: {\it QExecute} and {\it  QAlgorithmManager}.

\subsection{QExecute}

{\it QExecute} is the executor engine for {\it QuantumSolver}. It has the mission to authenticate against the IBM services, allowing to have access to the IBM systems (real hardware and online simulators), using an {\it “IBM Quantum Experience”} {\it API token} \cite{IBMQuantum} \cite{IBMQuantumAccount}. Also, {\it QuantumSolver} has a guest mode to avoid connect with the external IBM services and don't need a {\it token} and a user account in the {\it IBM Quantum} platform. In the guest mode, users can only run in local simulators ({\it ‘aer\_simulator’}), so to use real quantum system provided for IBM, users need to set up a {\it IBM Quantum} platform account. {\it QExecute} has different methods to visualize the list of available quantum {\it backends} and the selection of the chosen one to do the execution. Furthermore, {\it QExecute} is the component in charge of run the quantum circuits program.

\subsection{QAlgorithmManager}

{\it QAlgorithmManager} is the quantum algorithms manager of Quantum Solver. It will be the responsible to group and list all the available algorithms in the library, besides of selecting the one to run. Also, it allows to handle the arguments and parameters of the different algorithms and the information exchange between them with the main program.

\subsection{QAlgorithm}

{\it QAlgorithm} is the entity representation for a quantum algorithm in the Quantum Solver tool-set. It is a abstract class designed to use it like a algorithm template to add in a intuitive way new algorithms to the library. Any valid entity extended from the {\it QAlgorithm} entity will be a Quantum Solver algorithm ready to run in the tool-set. These entities who follow the {\it QAlgorithm} class template contain the relevant information about the specific algorithm: name, description, input parameters, output result, the explanation to analyze the quantum circuit result and the explanation to check and validate the input parameters introduced like a string. The main method of the entity is defined like a parameter generator for the quantum circuit mapped for the specific algorithm.

\section{Main program}

In the home screen, the Quantum Solver main program offers the different alternatives to play with it. You can run the guest mode or you can authenticate against your IBM account though the {\it API token}. In any case, a menu will appear with different options to visualize and select the different available {\it backends} and algorithms. When the user choose one of the algorithms to run in one of the backends, the program will request which parameters will be injected in the execution. When a {\it backend}, an algorithm and its parameters has been introduced, the program will show two different options: run one time or multiple times. On one side, the program allows to run the algorithm one time to get the result, showing a graphical representation of the input quantum circuit. On the other hand, it is possible to run the algorithm multiple times to observe the quantum behavior with a generated histogram in a mode called experimental.

\section{Interfaces}

\begin{figure}[tbh]
\centerline{
\includegraphics[width=8.5cm]{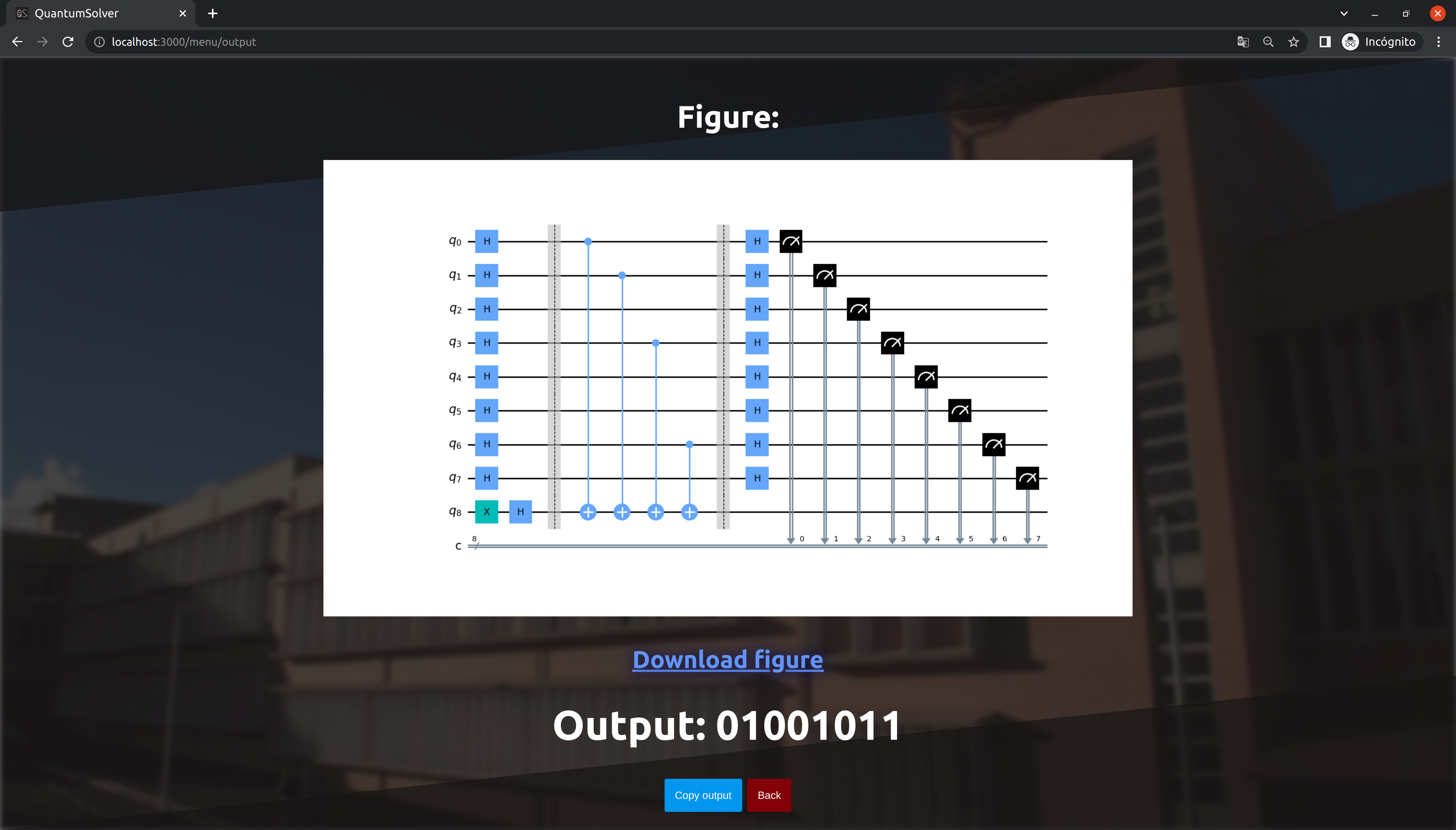}
}
\caption{{\it QuantumSolver} Web Interface }
\label{fig:run_web}
\end{figure}

{\it QuantumSolver} has a web platform version with a {\it backend} written in Python3, through the Flask {\it framework}. In the {\it frontend}, it has been written using TypeScript through React framework, HTML5 and CSS3. The web version offered by Quantum Solver (see Figure \ref{fig:run_web}) to run the quantum algorithms is the most intuitive for the general public. 

\begin{figure}[tbh]
\centerline{
\includegraphics[width=8.5cm]{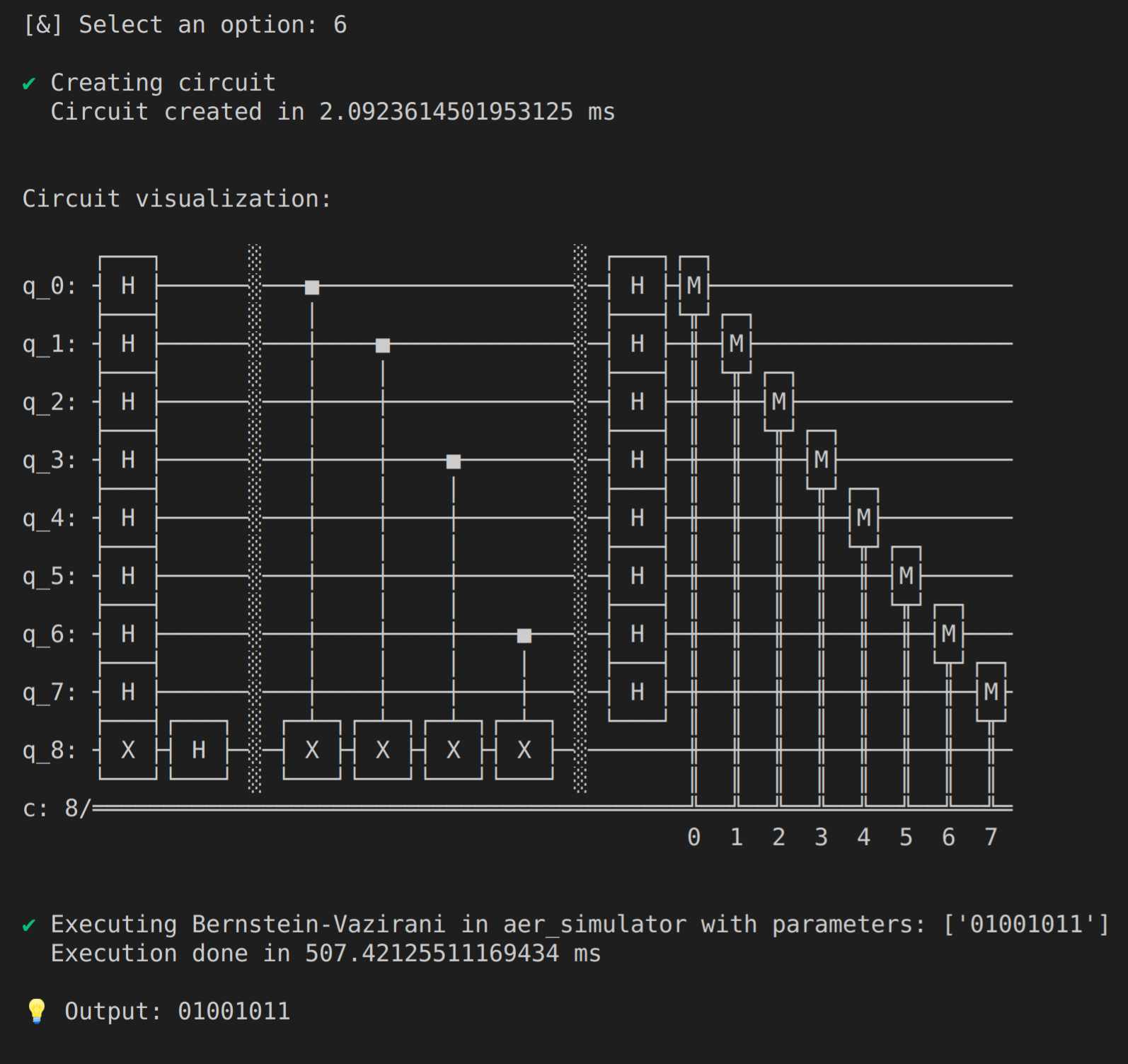}
}
\caption{{\it QuantumSolver} {\it CLI} (Command Line Interface)}
\label{fig:run_cli}
\end{figure}

The other one, the command line interface (CLI) (see Figure \ref{fig:run_cli}) is oriented to advanced users who are looking to integrate it in their development cycle. Both, web and CLI, they have the same features available to run and explore.

\section{Random numbers generator}

The {\it QRand} quantum algorithm implemented in Quantum Solver received like input parameter the natural number $(n)$ and it generates a quantum circuit ready to run and get like result a random number between $0$ y $2^n - 1$. Its operation is based on the initialization of $n$ qubits, by default to $\ket{0}$ state. Then, the insertion of a Hadamard \cite{LearnQiskit} quantum logic gate to every qubit to generate a superposition state where the probability to measure $0$ or $1$ (ver Ec. (\ref{eq:qrand})) is the same. Finally, the result measurement, collapsing each qubit in a random state and mapping that like a binary number.

\begin{equation}
|00\dots 0\rangle \xrightarrow{H^{\otimes n}} \frac{1}{\sqrt{2^n}} \sum_{x\in \{0,1\}^n} |x\rangle
\label{eq:qrand}
\end{equation}

\section{Bernstein-Vazirani Algorithm}

The Bernstein-Vazirani \cite{BernsteinVazirani} algorithm included in Quantum Solver receives like input parameter a key composed by an arbitrary binary string with $n$ length which it is used to code an oracle representing the information of the key. In specific, against a candidate string, the information returned by the algorithm is the confirmation about if the $1$ number of coincidences between the key strings a the candidate string is odd or even (See equation (\ref{eq:oracle_bernstein_vazirani})).

\begin{equation}
f_s(x) = (s * x) (mod~2) 
\label{eq:oracle_bernstein_vazirani}
\end{equation}

The mathematical formulation represented in (\ref{eq:oracle_bernstein_vazirani}) shows how the oracle perform the binary product between the bits pairs of the key to guess and the candidate key. Then it applies a $XOR$ gate or mod 2 addition to the other $n$ bits. 
In a classical world, this problem can be solved using $n$ queries against the oracle.

\begin{equation}
\begin{cases}
f_s(100 \ldots 0) = s_{0} \\
f_s(010 \ldots 0) = s_{1} \\
f_s(001 \ldots 0) = s_{2} \\
\ldots \\
f_s(000 \ldots 1) = s_{n - 1}
\end{cases}
\label{eq:bernstein_vazirani_classic_solution}
\end{equation}

In the quantum case, it only need a query against the oracle, returning the right key with a 100\% of probabilities (without consider the possible errors generated by noise in the {\it system}).

The implementation of the algorithm \cite{QiskitBernsteinVazirani} requires $n + 1$ qubits. From these qubits, $n$ are for code the input (built for $n$ qubits with a $\ket{0}$ state value). The other qubit is additional for the quantum oracle output (initialized with a $\ket{1}$ value state obtained to apply the quantum gate $X$ to one qubit with a $\ket{0}$ value state by default). Futhermore, it will be applied Hadamard quantum gates to $n + 1$ qubits, previous to the oracle and after to the oracle, excepting in the output which it will not be required because will not be measured.

Internally, the quantum oracle should be implemented through {\it CNOT} gates controlling the qubits which represent the bits for the key with value state to $1$ and the goal is the quantum oracle output. This gate {\it CNOT} is the equivalent to the classic {\it XOR}.

The oracle characterization applied over the candidate string $x$, it is showed in the formula (\ref{eq:quantum_oracle_bernstein_vazirani}).

\begin{equation}
|x \rangle \xrightarrow{f_s} (-1)^{s\cdot x} |x \rangle
\label{eq:quantum_oracle_bernstein_vazirani}
\end{equation}

The performed transformation for the $n$ qubits which codify the input $\ket{00\dots0}$ using the Hadamard quantum gates in the initialization process is showed in the formula (\ref{eq:initialization}).

\begin{equation}
|00\dots 0\rangle \xrightarrow{H^{\otimes n}} \frac{1}{\sqrt{2^n}} \sum_{x\in \{0,1\}^n} |x\rangle
\label{eq:initialization}
\end{equation}

The quantum oracle application is showed in the formula (\ref{eq:applying_oracle}).

\begin{equation}
\frac{1}{\sqrt{2^n}} \sum_{x\in \{0,1\}^n} |x\rangle \xrightarrow{f_a} \frac{1}{\sqrt{2^n}} \sum_{x\in \{0,1\}^n} (-1)^{a\cdot x}|x\rangle
\label{eq:applying_oracle}
\end{equation}

The final step which consist to apply for each qubit a Hadamard gate, it is showed in the formula (\ref{eq:applying_last_hadamards}).

\begin{equation}
\frac{1}{\sqrt{2^n}} \sum_{x\in \{0,1\}^n} (-1)^{a\cdot x}|x\rangle \xrightarrow{H^{\otimes n}} |a\rangle
\label{eq:applying_last_hadamards}
\end{equation}

It observes the encrypted key in the oracle is the result to apply the described operations in this article. 

\section{BB84 Protocol}
The cryptography BB84 \cite{BB84Protocol} protocol for the quantum key distributions included in the Quantum Solver tool-set.

\subsection{Entities}
It has been implemented a main entity ({\it Participant}) and its derivative entities ({\it Sender} and {\it Receiver}).
It has been supposed that a class instance {\it Sender} wants to communicate with the other class instance {\it Receiver}, with a secret communication thanks to the BB84 protocol which it allows a 
one-time pad generated shared between the Sender and the Receiver. The quantum channel simulation is described through quantum circuits using Qiskit \cite{QiskitBB84}. The base class {\it Participant} contains the methods to generate and draw the values, axis, keys and one-time pads, etc.

\subsection{Basis}
The Sender entity chooses, in a random way, values to encrypt each qubit to transmit and the axis to code them. The results obtained are the probabilities showed in the Table \ref{tab:valor_eje_circuito}.

\begin{table}[htb]
\centering
\caption{Value-Axis possibilities and the associated quantum circuits}
\label{tab:valor_eje_circuito}
\begin{tabular}{c c c}
\hline
\hline
 Value & Axis & Circuit \\
\hline

 $\ket{0}$ & Z & \parbox[c]{0.35\linewidth}{\scalebox{1.0}{
\Qcircuit @C=1.0em @R=1.0em @!R { \\
                \nghost{{q} :  } & \lstick{{q} :  } & \qw & \qw\\
\\ }}} \\
 $\ket{0}$ & X & \parbox[c]{0.35\linewidth}{\scalebox{1.0}{
\Qcircuit @C=1.0em @R=0.2em @!R { \\
                \nghost{{q} :  } & \lstick{{q} :  } & \gate{\mathrm{H}} & \qw & \qw\\
\\ }}} \\
 $\ket{1}$ & Z & \parbox[c]{0.35\linewidth}{\scalebox{1.0}{
\Qcircuit @C=1.0em @R=0.2em @!R { \\
                \nghost{{q} :  } & \lstick{{q} :  } & \gate{\mathrm{X}} & \qw & \qw\\
\\ }}} \\
 $\ket{1}$ & X & \parbox[c]{0.35\linewidth}{\scalebox{1.0}{
\Qcircuit @C=1.0em @R=0.2em @!R { \\
                \nghost{{q} :  } & \lstick{{q} :  } & \gate{\mathrm{X}} & \gate{\mathrm{H}} & \qw & \qw\\
\\ }}} \\

\hline
\hline
\end{tabular}
\end{table}

Coming up next, the Receiver entity get the quantum circuits (each one represents a qubit) and in a random way it choose axis where measure it. Around 50\% of the qubits will be measured with right values, it means in the same axis than the Sender entity. The other 50\% will be discarded because the axis measured is the wrong one, so it will be a 50\% of probabilities to send the correct encrypted value. It provokes a lost in the information sent. To know which values are discarded, both entities make public the axis to measure the qubits, because it doesn't exist a real risk with it. In this way, they can avoid the values where the axis are not the same.

The output values after clean the wrong ones, they can be considered like the generated key, but it needs to verify the security before of that. It exists a incoherence between the key sent by the Sender and the key received by the Receiver with the right axis. This is that around the 50\% of the qubits measured in the wrong axis by a hacker \cite{PaperBB84}. In the table \ref{tab:posibilidades_bb84} 4 cases about measurements in the qubits are shown, all of them with a 25\% of probabilities to happen.

\begin{table*}[htb]
\centering
\caption{Cases about measurments in a qubit in the verification step for BB84}
\begin{tabular}{p{.2\textwidth} p{.2\textwidth} p{.6\textwidth}}
\hline
\hline
 Measurement by trust Sender and Receiver & Measurement by in the middle Attack & Conclusions about the qubit \\
\hline

 Different Axis & Sender Axis & The value is discarded although the intermediate Receiver measured with a 100\% of probabilities to get the right value \\
 \hdashline
 Different Axis & Receiver Axis & The values is discarded although the intermediate Receiver measured with a 50\% of probabilities to get the right value (total uncertainty value) \\
 \hdashline
 Same Axis & Same Axis for Sender and Receiver & It has been intercepted by the hacker without cancel the protocol, with a 100\% of probabilities to get the right value \\
 \hdashline
 Same Axis & Opposite Axis for Sender and Receiver & It exists a 50\% of probabilities to abort the protocol, in case of the Sender collapse with the opposite value emitted by the Sender \\

\hline
\hline
\end{tabular}
\label{tab:posibilidades_bb84}
\end{table*}

In the last item in the Table \ref{tab:posibilidades_bb84}, it has been observed with a probability of 50\% an cancellation of the protocol thanks to detect the insecure communication. Taking into account one of the four cases has a 25\% of probabilities, when a unique qubit is sent the protocol is aborted with a 12.5\% of $p_{detect}$ probability. When $n$ qubits are sent, because the against to any kind of a detection about a malicious behavior, the probability to cancel the protocol is related with not detect any malicious behavior from the formula (\ref{eq:abortar_protocolo_prob}). 

\begin{equation}
1 - (1 - p_{detect})^n = 1 - \left(1 - \frac{1}{4} * \frac{1}{2}\right) ^n = 1 - \left( \frac{7}{8} \right) ^n
\label{eq:abortar_protocolo_prob}
\end{equation}

This value is also obtained after to apply the Bayes Theorem with the probabilities to not detect any attack against a qubit not discarded (3/4) and on the qubit discarded (1)

To verify the protocol result, the Receiver publish the half of the obtained key to allow to the Sender compares it with the half of its key. If both are the same, the protocol execution is secure and the other half of the key can be used like shared secret key. In other cases, it supposes that any hacker did an malicious action against the protocol through a {\it eavesdropping} attack, intercepting the sent qubits and measure them before to reach the legit and final Receiver.

\subsection{Program}

The {\it QuantumSolver} library allows the execution of a implementation for the BB84 protocol. To run the program, a menu is deployed in the screen to facilitate the visualization and selection of the {\it backend} available. After to choose the desired backend, two possible options appears for the user. On one side, the user can run the program one time and visualize the trace between the different participants in the communication. For that it is necessary specify a string like a message and a value between $0$ and $1$ like interception density (it is the hacker probability to measure a qubit). On the other side, the user can run the protocol several times to show a heating map where light colors represent the cases when the communications has been considered like secure, and dark colors represents when the communications has been intercepted by any malicious user before to reach the message the legit receiver. For this last mode, considered experimental mode, it have to specific different conditions and assumptions: the max length for the string in number of bits (which it defines the axis $x$ in the heating map) defined to get positives integer values up to the this max; the value for the interception density {\it step} (which it defines the axis $y$ in the heating map) defined to get values between $0$ and $1$; and the number of iterations per each generated problem instance.

\begin{figure}[tb]
\centerline{
\includegraphics[width=8.3cm]{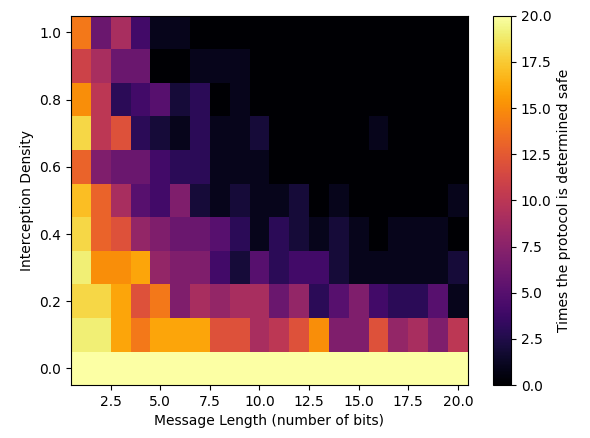}
}
\caption{First Example about heating map for BB84}
\label{fig:primerMapaDeCalorBB84}
\end{figure}

\begin{figure}[tb]
\centerline{
\includegraphics[width=8.3cm]{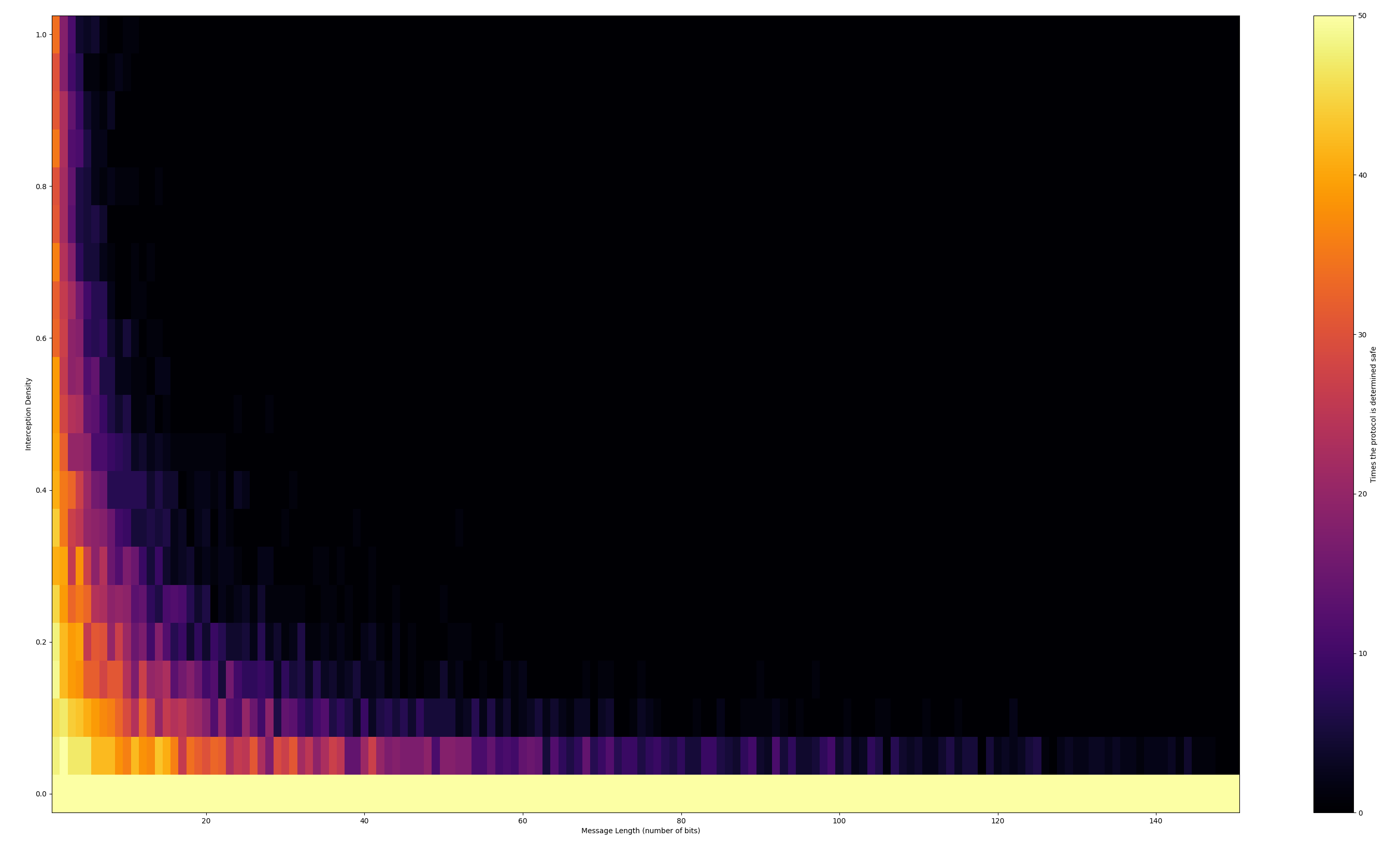}
}
\caption{Second Example about heating map for BB84}
\label{fig:segundoMapaDeCalorBB84}
\end{figure}

 The Figures \ref{fig:primerMapaDeCalorBB84} and \ref{fig:segundoMapaDeCalorBB84} show the two examples about heating map generated, regarding to the parameters shown in the Table \ref{tab:mapas_de_color}. 

\begin{table}[ht]
\centering
\caption{Relevant parameters for the heating maps}
\label{tab:mapas_de_color}
\begin{tabular}{lllll}
\hline
\hline
 Figure & Message Max length & Interception density {\it Step} & Iteration & Execution time\\
\hline

 \ref{fig:primerMapaDeCalorBB84} & 20 bits & 0.1 & 20 & 5 minutes \\
 \ref{fig:segundoMapaDeCalorBB84} & 150 bits & 0.05 & 50 & 960 minutes \\

\hline
\hline
\end{tabular}
\end{table}

\section{Conclusions}

The implementation introduced in this paper for the quantum tool-set Quantum Solver allows in a easy way , the execution of several quantum algorithms in quantum {\it systems} (real and simulators devices) provided by IBM. Also, it offers a simple architecture about Entities which it makes easier the addition of new algorithms to the designed library. In specific, the main goal of the project is reuse the {\it tool-set} from a developer point of view allowing increase the number of algorithms available in the library. It includes new algorithms like B92, E91 or the Grover Algorithm.

\section*{Acknowledgment}

This investigation is supported by the RTI2018-097263-B-I00 project thanks to the financing from Ministerio de Ciencia, Innovación y Universidades, the Agencia Estatal de Investigación and the Fondo Europeo de Desarrollo Regional, and the la Cátedra de Ciberseguridad Binter-Universidad de La Laguna.

%

\end{document}